# Emergence of Giant Magnetic Chirality during Dimensionality Crossover of Magnetic Materials

Dae-Yun Kim[1,2,†], Yun-Seok Nam[1,3,†], Younghak Kim[4,†], Kyoung-Whan Kim[5], Gyungchoon Go[6], Seong-Hyub Lee[1,3], Joon Moon[3], Jun-Young Chang[1,3], Ah-Yeon Lee[7], Seung-Young Park[7], Byoung-Chul Min[1], Kyung-Jin Lee[6], Hyunsoo Yang[2\*], Duck-Ho Kim[1\*], and Sug-Bong Choe[3\*]

[1]Center for Spintronics, Korea Institute of Science and Technology (KIST), Seoul 02792, Republic of Korea

[2]Department of Electrical and Computer Engineering, National University of Singapore, Singapore 117583, Singapore

[3]Department of Physics and Institute of Applied Physics, Seoul National University, Seoul 08826, Republic of Korea

[4]Pohang Accelerator Laboratory, Pohang University of Science and Technology (POSTECH), Pohang, Gyeongbuk 37673, Republic of Korea

[5]Department of Physics, Yonsei University, Seoul 03722, Republic of Korea

[6]Department of Physics, Korea Advanced Institute of Science and Technology (KAIST), Daejeon 34141, Republic of Korea

[7]Center for Scientific Instrumentation, Korea Basic Science Institute (KBSI), Daejeon 34133, Republic of Korea

[†]These authors are contributed equally to this work.

[\*]e-mail: eleyang@nus.edu.sg (H. Yang), uzes@kist.re.kr (D.-H. Kim), sugbong@snu.ac.kr (S.-B. Choe)

Chirality, an intrinsic preference for a specific handedness, is a fundamental characteristic observed in nature. In magnetism, magnetic chirality arises from the anti-symmetric Dzyaloshinskii-Moriya interaction in competition with the symmetric Heisenberg exchange interaction. Traditionally, the anti-symmetric interaction has been considered minor relative to the symmetric interaction. In this study, we demonstrate an observation of giant magnetic chirality during the dimensionality crossover of magnetic materials from three-dimensional to two-dimensional. The ratio between the anti-symmetric and symmetric interactions exhibits a reversal in their dominance over this crossover, overturning the traditional consideration. This observation is validated theoretically using a non-local interaction model and tight-binding calculation with distinct pairing schemes for each exchange interaction throughout the crossover. Additional experiments investigating the asphericity of orbital moments corroborate the robustness of our findings. Our findings highlight the critical role of dimensionality in shaping magnetic chirality and offer strategies for engineering chiral magnet states with unprecedented strength, desired for the design of spintronic materials.

Symmetry, a fundamental principle in nature, underpins diverse phenomena across physics, mathematics, and other scientific disciplines[1–9]. It implies the invariance of certain physical properties under transformations such as spatial translation, time reversal, or rotation[10], forming the cornerstone of universal physical laws[4–9]. However, the breaking of symmetry often results in novel and intriguing physical phenomena[7–9, 11–14]. Particularly, the breaking of inversion symmetry generates chirality, endowing systems with a distinct preference for a particular handedness in their structures. Chirality, a concept with far-reaching implications, has been extensively studied across scientific fields for its role in shaping diverse natural and engineered systems[3,7–9,11–14].

In magnetism, chirality is an especially critical area of investigation owing to its fundamental role in modern spintronics[7–9,11–14]. The breaking of inversion symmetry in magnetic materials results in an anti-symmetric exchange interaction known as the Dzyaloshinskii-Moriya interaction (DMI)[8,9], which governs both the magnetization dynamics and static spin configurations of magnetic systems[11–20]. Notably, DMI is crucial in generating chiral magnetic texture and enabling effective operation for next-generation spintronic devices, including magnetic random-access memory and magnetic racetrack memories. Thus, enhancing magnetic chirality is a key focus for improving the efficiency and performance of these devices, prompting efforts to maximize its strength.

Magnetic chirality can be quantitatively characterized by the ratio of the anti-symmetric exchange interaction (DMI) to the symmetric exchange interaction (Heisenberg exchange), which can be expressed as[21]:

$$\chi \equiv D_{\text{ex}}/A_{\text{ex}}, \quad (1)$$

Here, $D_{\text{ex}}$ and $A_{\text{ex}}$ represent the DMI constant[8,9] and Heisenberg exchange stiffness,

respectively, whose precise definitions to match the unit are shown below. While the symmetric exchange favors uniform spin configurations, the anti-symmetric exchange interaction induces non-uniform, spiral-like spin textures with a preferred handedness. Traditionally, the DMI has been widely regarded as significantly weaker than the Heisenberg exchange interaction, and often considered a perturbative effect[8,9,11,21–25]. Both theoretical and experimental studies have consistently shown that $\chi$ is typically much smaller than unity (i.e., $\chi \cong 4.7\pm0.7\%$),[22] which imposes a fundamental limitation on efficiency and performance of spintronic devices, including implementing skyrmion racetrack memory, enhancing magnetic domain-wall (DW) speed, enabling ultra-energy-efficient magnetization switching, and enabling chiral damping manipulation. Overcoming this constraint remains a key challenge in advancing spintronic technologies and fundamental physics.

To date, research has predominantly concentrated on the DMI, with limited attention given to the ratio $D_{ex}/A_{ex}$, largely due to the experimental challenges in quantifying $A_{ex}$, particularly in magnetic atomically-thin magnetic films. In this study, we systematically investigated the strength of magnetic chirality, denoted as $\chi$, by reducing the dimensionality[26–37] in magnetic materials. We identify the emergence of giant magnetic chirality during the dimensional crossover from three-dimensional (3D) to two-dimensional (2D) systems, with $\chi$ exceeding unity (i.e., $\chi \cong 150\%$ of maximal value) in this regime. Our findings reveal that the underlying mechanism for this giant magnetic chirality is the distinct non-local pairing of symmetric and anti-symmetric exchange interactions, which varies with the dimensionality of the magnetic system. Furthermore, additional experimental investigations, specifically examining the asphericity of orbital moments[38,39], confirm the consistency of our findings with both experimental observations and theoretical predictions. This work establishes dimensionality as a fundamental factor in the manifestation of magnetic chirality, offering a new framework for engineering robust chiral magnetic states.

We aim to postulate the potential transition of exchange interactions during the dimensional crossover of magnetic materials. The Heisenberg exchange interaction arises from the interaction between neighboring atoms within the magnetic layer, with both the source and target represented by atoms from the magnetic layer, as depicted in Fig. 1**a**. Contrastingly, the DMI, which occurs at the interface between the magnetic and non-magnetic layers, primarily originates from interactions involving conduction electrons traveling along the path connecting adjacent atoms from the non-magnetic and magnetic layers as shown in Fig. 1**b**. In this case, atoms from the non-magnetic layer act as the source, and atoms from the magnetic layer act as the target, as illustrated in Fig. 1**b**. From an intuitive perspective, the number of *atomic pairs* from the magnetic layer determines the Heisenberg exchange interaction, whereas the number of *atomic sites* within the magnetic layer is associated with the DMI. Accordingly, the DMI unit pair can be defined as a pair of atoms at the interface of the non-magnetic and magnetic layers, while the Heisenberg exchange interaction unit pair involves two atoms from the magnetic layer.

A reduction in the magnetic layer thickness decreases the number of atomic sites within the magnetic layer, while the number of atoms from the non-magnetic layer (acting as the source for DMI) remains constant. However, the number of atoms from the magnetic layer (acting as the target) also decreases, leading to a decline in both the atomic pairs involved in the Heisenberg exchange interaction and the atomic sites participating in the DMI. Notably, the reduction in the number of atoms from the magnetic layer is expected to cause a more rapid decline in the number of atomic pairs contributing to the Heisenberg exchange interaction compared to the decline in atomic sites associated with the DMI. For a straightforward conceptual understanding, the schematic illustration is shown in Fig. 1**c**. This unexpected outcome highlights the critical role of DMI under reduced thickness conditions, challenging initial assumptions based on conventional exchange interactions.

To examine this unexpected postulation, various series of magnetic thin films with perpendicular magnetic anisotropy were fabricated for this study. Their structure was Si substrate/ SiO$_2$ (100 nm)/ Ta (5.0 nm)/ Pt (4.0 nm)/ Co ($t_{\text{eff}}$ nm)/ X (2.5 nm)/ Pt (1.5 nm), where $t_{\text{eff}}$ represents the effective thickness of the magnetic layer, adjusted to account for the magnetic dead layer (see Supplementary Information Note S1). The capping layer X corresponds to four different non-magnetic materials (X = Ta, Ti, Ru, and Pd), designed to tune the anti-symmetric exchange interactions. The thickness of the magnetic layer was varied over a wide range, from a few atomic monolayers to several nanometers, fully encompassing the dimensional crossover regime between 3D and 2D. The top 1.5-nm Pt layer served as a capping layer to prevent oxidation of X.

Both $A_{\text{ex}}$ and $D_{\text{ex}}$ were investigated in the films by measuring the spin-torques acting on magnetic DWs[40,41] (see Supplementary Information Note S1). Figure 2**a** illustrates the spin torque measurement results for one of the samples, where $\varepsilon_{\text{ST}}$ represents the spin torque acting on DWs per unit current density and $\mu_0 H_x$ denotes the in-plane magnetic field applied to DWs to control their chirality. Guided by solid curves and arrows, one can quantify DW anisotropy field $\mu_0 H_S$ and DMI-induced field $\mu_0 H_{\text{DMI}}$. $A_{\text{ex}}$ can be measured by using the DW anisotropy field $\mu_0 H_S$ in a thin ferromagnet[41] and $D_{\text{ex}}$ is determined based on $\mu_0 H_{\text{DMI}}$ using the relation $D_{\text{ex}} = D_0 t_{\text{eff}} = (\lambda M_S \mu_0 H_{\text{DMI}}) t_{\text{eff}}$, where $\lambda$ and $M_S$ represent the width of the DW and the saturation magnetization (see Supplementary Information Note S1 for the detailed relation). Parameter $D_0$ represents the conventional DMI constant [J/m$^2$], which is typically known to be inversely proportional to $t_{\text{eff}}$. $A_{\text{ex}}$ and $D_{\text{ex}}$ quantify the symmetric and anti-symmetric exchange interactions in the identical unit [J/m], and therefore a direct comparison between them revealed the dominance between the two exchange interactions.

Figure 2**b** plots $A_{ex}$ and $D_{ex}$ as a function of $t_{eff}$ for X = Ta. Both $A_{ex}$ and $D_{ex}$ decreases with decreasing $t_{eff}$ (within 3–4 atomic monolayers of the ferromagnetic layer). The crossover point in the figure indicates a considerably more rapid decrease for $A_{ex}$ than for $D_{ex}$ when the system is scaled down to the a few atomic monolayer. Figure 2**c** presents analogous outcomes that hold true irrespective of the X materials (X = Ti, Ru, and Pd) involved. These findings indicate that, in the quasi 2D regime, $D_{ex}$ approaches a magnitude comparable to that of $A_{ex}$, and in some cases, it even surpasses the magnitude of $A_{ex}$. This observation challenges the prevailing knowledges from two distinct aspects: 1), the current understanding of the energetic scale of anti-symmetric exchange interaction as being only a small fraction of the symmetric exchange interaction energy scale[8,9,11,21–24]; and 2), the long-standing assumption of a direct proportionality between symmetric and anti-symmetric exchange interactions, across various systems, including bulk magnetic oxides, metallic spin glasses with magnetic impurities, and metallic ferromagnets[9,21,42–44].

Figure 2**d** plots the ratio $\chi \equiv D_{ex}/A_{ex}$ with respect to $t_{eff}$ for X = Ta, Ti, Ru, and Pd, respectively. There are two interesting features in this figure; First, a drastic enhancement of $D_{ex}/A_{ex}$ by decreasing $t_{eff}$ is observed in the quasi 2D regime. This is in stark contrast to the common belief[22] that $D_0/A_{ex}$ is inversely proportional to $t_{eff}$ such that $D_{ex}/A_{ex}$ is expected to be independent of $t_{eff}$. Second, a large $D_{ex}/A_{ex}$ (>100%) is observed. The highest $D_{ex}/A_{ex}$ is ~150%, which is orders of magnitude larger than the previously reported values for ferromagnets (4.7±0.7%, as depicted by the star symbol in Fig. 2**d** from the previous study[22]).

Here, we propose a non-local interaction model to establish a comprehensive theoretical framework for understanding the dependence of $A_{ex}$ and $D_{ex}$ on magnetic layer thickness, extending beyond the limitations of the traditional local approximation. The conceptual framework of the Heisenberg exchange interaction is linked to the volume of the magnetic layer as shown in Fig. 3**a**. The exchange energy between the local magnetic moments at sites *i* and *j* can be expressed as the product of a decaying function and the Heisenberg exchange energy contributed by each atom. We obtain the average $A_{ex}$ by summing these exchange energies. In the traditional local approximation model, the decaying function is modelled as a muffin-tin function, which is zero for distances between the sites of two atoms that are greater than the lattice constant $a$, and constant for distances smaller than $a$. The latter results in $A_{ex}$ being independent of the volume, and thus the thickness. However, in ultrathin ferromagnets with thicknesses comparable to the lattice constant, the muffin-tin approximation is modified with an exponentially decaying function with the characteristic length $\xi_{EX}$. We analytically derive the relationship $A_{ex} = \frac{\tilde{A}}{2}\left[(2 + e^{-t/\xi_{EX}}) - 3(1 - e^{-t/\xi_{EX}})\frac{\xi_{EX}}{t}\right]$, by employing a continuum model for the site indices. In this relationship, $t$ is the thickness of the magnetic layer and $\tilde{A} = \lim_{t \to \infty} A_{ex}$ represents the bulk exchange stiffness (see Supplementary Information Note S2 for details). Figure 3**b** illustrates the conceptual framework of the DMI relative to the volume of the magnetic layers, where the DMI between the local magnetic moments (*i* and *j* site atoms) and the heavy metal atoms (*k* site atom) can be expressed as the product of a decaying function and the DMI energy contributed by each atom. We derive the following analytical relationship using the same continuum model: $D_{ex} = \tilde{D}_{ex}\left[1 - e^{-t/\xi_{DM}}\left(1 + \frac{t}{2\xi_{DM}}\right)\right]^2$, where $\tilde{D}_{ex}$ is determined by $\lim_{t \to \infty} D_0 t$ and $\xi_{DM}$ represents the characteristic length of the decay of the DMI (see Supplementary Information Note S2 for details).

Figures 3**c** and 3**d** plots $A_{ex}$ and $D_{ex}$ with respect to the magnetic layer thickness scaled by the characteristic length of the decay of the exchange interactions ($\xi_{EX}$ and $\xi_{DM}$, respectively). $A_{ex}$ decreases more sharply compared to $D_{ex}$ when the ferromagnetic thickness decreases, which can be explained intuitively by the number of atomic pairs linked to $A_{ex}$, with the number of atomic sites being associated with $D_{ex}$. From our theoretical model, we can obtain the ratio between two exchange interactions, $\chi^{Model} \equiv D_{ex}/A_{ex}$, which is expressed by

$$\chi^{Model} = \left(\frac{\widetilde{D}_{ex}}{\widetilde{A}}\right) \frac{2\left[1-e^{-t/\xi_{DM}}\left(1+\frac{t}{2\xi_{DM}}\right)\right]^2}{(2+e^{-t/\xi_{EX}})-3(1-e^{-t/\xi_{EX}})\frac{\xi_{EX}}{t}}. \tag{2}$$

Figure 3**e** shows the normalized $\chi^{Model}$, the ratio between $\chi^{Model}$ and $\chi^{Model}_{Bulk}(\equiv \widetilde{D}_{ex}/\widetilde{A})$ as a function of $t/\xi_{EX}$ for $\xi_{EX}/\xi_{DM} = 7$ from Eq. (2). The figure clearly shows that $\chi^{Model}/\chi^{Model}_{Bulk}$ increases at ultranarrow magnetic layer thickness, implying that the magnetic chirality is boosted extraordinarily at the atomically thin magnets, as observed experimentally.

To validate our analytic results on the dimensionality dependence of magnetic chirality, we perform a numerical calculation based on a tight-binding (TB) model. In this calculation, the Heisenberg exchange interaction is captured by electron hopping processes in the bulk square lattice of the magnetic layer and the DMI is generated via hybridization between atoms of top non-magnetic and magnetic layers[44] (see Supplementary Information Note S3 for details). Figure 3**f** shows the numerical calculation result of the magnetic chirality $\chi^{TB}$ normalized by the bulk value denoted by $\chi^{TB}_{Bulk}$ with respect to $t$ using a minimal TB model[44]. Here $\chi^{TB}_{Bulk}$ represents the ratio of the two exchange interactions for a large ferromagnet in the TB calculation. This result is consistent with our experimental findings (Fig. 2**d**) and analytic model (Fig. 3**e**). Therefore, the pivotal ramification of our results suggests that the anti-symmetric exchange does not constitute a minor fraction of the symmetric exchange. Further, a noteworthy departure from conventional expectations is observed because dominance between two exchange interactions is reversed upon reducing the thickness of the ferromagnetic layer to a mere few atomic monolayers.

The ratio of symmetric to anti-symmetric exchange interactions, a key measure of magnetic chirality, has long been recognized as a fundamental parameter in magnetism. Previously, the Rashba constant was intricately linked to this ratio[21] and the asphericity of orbital moments[38,39,45]. According to previous studies[38,39], the Rashba constant is simply influenced by the atomic spin-orbit coupling (SOC) constant $\xi$ of the interfacial states and the surface potential gradient. The latter is directly proportional to the ratio between the in-plane and out-of-plane orbital moments ($m_\parallel$ and $m_\perp$). Given that orbital moment asphericity is directly proportional to the exchange interaction ratio, we performed orbital moment measurements to substantiate our findings. Consequently, the magnetic chirality from the anisotropy of orbital moments, denoted by $\tilde{\chi}$, is determined to be approximately equal to $\xi$ multiplied by the ratio of out-of-plane to in-plane orbital moments ($m_\perp/m_\parallel$)[39]: $\tilde{\chi} \equiv \xi(m_\perp/m_\parallel)$.

Figure 4**a** presents X-ray absorption spectroscopy (XAS) spectra (solid black line) which are averaged by right- and left-handed circularly polarized X-ray spectra at the Co L$_{2,3}$ edges obtained at $0°$ for X = Ti and $t_{\text{eff}} = 0.6$ nm. The solid green line indicates integrated XAS spectra and $A_{\text{total}}$ is determined. Figure 4**b** shows X-ray magnetic circular dichroism (XMCD) spectra (solid black line) corresponding to the difference between right- and left-handed circularly polarized X-ray spectra at the Co L$_{2,3}$ edges obtained at $0°$ for X = Ti and $t_{\text{eff}} = 0.6$ nm. The solid red and blue lines denote integrated XMCD spectra for each energy

ranges for determining $\Delta A_{L3}$ and $\Delta A_{L2}$. By following a sum-rule calculation, one can estimate $m_\parallel$ and $m_\perp$ (see Supplementary Information Note S4 for details).

Figure 4**c** shows the magnetic chirality scaled by the SOC constant expressed as $\tilde{\chi}/\xi$ ($\equiv m_\perp/m_\parallel$) from the orbital moments' measurement as a function of $t_{\text{eff}}$ for X = Ta, Ti, Ru, and Pd, respectively. We observe a clear thickness dependence of $\tilde{\chi}/\xi$, which is consistent with the thickness dependence of $\chi$ shown in Fig. 2**d**. The dimensionality transition thicknesses $t_C$ can be determined using $\chi$ (see Fig. 2**d**) and $\tilde{\chi}/\xi$ (see Fig. 4**c**) curves, which are ~1.09 ± 0.17 nm and 1.15 nm, respectively; these values coincide with each other, as also observed in the theoretical analysis (see Figs. 3**e** and 3**f**). This correspondence supports our finding of the dimensionality transition of the magnetic chirality.

The significance of our findings can be summarized in two key aspects. First, the unexpected relative strength of the magnetic chirality can lead to the emergence of a helimagnetic ground state in ultrathin magnets, possibly instigating robust stabilization of the topological spin textures because the ratio between exchange interactions measures the pitch of the chiral modulations in the ordered helical states[11,14,24,25]. Second, the exploration of low-dimensional materials provides an avenue for unveiling new realms of physics, specifically regarding emerging nonlocal pairing exchange interactions or the strongly interatomic-orbital overlap originating from the surface observed within 2D magnets[35,46].

In summary, we observe the emergence of a giant magnetic chirality by reducing the dimensionality of ferromagnet from 3D to 2D. In the quasi-2D regime, the anti-symmetric exchange interaction (DMI) surpasses the symmetric exchange interaction (Heisenberg exchange), exceeding 100%, which challenges the conventional perturbative energy threshold. Theoretical supports are provided by validating a non-local interaction model and tight-binding calculation. These results highlight the role of magnetic volume in shaping exchange interactions, providing a new approach for enhancing magnetic chirality in spintronic applications.

**Figure Captions**

**Figure 1. Diagram Depicting the Dimensional Transition in the Interplay of Bulk and Interface Energies. a**, The Heisenberg exchange interaction arises from the interaction between neighboring atoms within the magnetic layer, with both the source and target represented by atoms from the magnetic layer. **b**, DMI, which occurs at the interface between the magnetic and non-magnetic layers, primarily originates from interactions involving conduction electrons traveling along the path connecting adjacent atoms from the non-magnetic and magnetic layers. **c**, Conceptual schematics for the reduction in the number of atoms from the magnetic layer contributing to the Heisenberg exchange interaction and to the DMI across a transition from three-dimensional (3D) to two-dimensional (2D) scenarios.

**Figure 2. Thickness Dependence of Exchange Interactions and the Magnetic Chirality. a**, plot of the DW chirality and the spin torque efficiency $\varepsilon_{ST}$ as a function of the in-plane magnetic field $\mu_0 H_x$. Red and blue arrows represent $\mu_0 H_S$ and $\mu_0 H_{DMI}$, respectively. Vertical dashed lines indicate where the DW chirality is of Bloch-type and Neel-type. **b**, Plot of the Heisenberg exchange stiffness constant ($A_{ex}$) and the Dzyaloshinskii-Moriya interaction (DMI) constant at the interface ($D_{ex}$) as a function of the effective thickness of the Co layer, rescaled with consideration of magnetic dead layer ($t_{eff}$) for X=Ta. The dashed dot represents a guideline for the eye. **c**, Plot of $A_{ex}$ and $D_{ex}$ as a function of $t_{eff}$ for X=Ti, Ru, and Pd. The dashed dots represent guidelines for the eye. **d**, Plot of $t_{eff}$ dependent magnetic chirality $\chi \equiv D_{ex}/A_{ex}$ for X = Ta, Ti, Ru, and Pd, respectively. The colored dashed lines indicate the best linear fit for each X.

**Figure 3. Theoretical Methods for Modelling of Exchange Interactions Utilizing a Non-local Interaction Analytical Model and a Tight-binding Calculation. a-b**, Figurative depiction demonstrating the Heisenberg exchange interaction, emanating from the interplay between adjacent atoms within the ferromagnetic material, wherein the atoms themselves serve as both source and target. Illustrative representation of the DMI occurring at the interface of a bilayer composed of a ferromagnetic material and a heavy metal. It primarily stems from the interaction between conduction electrons traversing the pathway connecting the heavy metal and ferromagnetic atoms. Here the heavy metal atom assumes the role of the electron's source, while the ferromagnetic atom becomes the target, as perceived from the electron's perspective. **c-d**, Plot of $A_{ex}$ and $D_{ex}$ with respect to ferromagnetic thickness scaled by the characteristic length of the decay of the exchange interactions ($\xi_{EX}$ and $\xi_{DM}$, respectively). **e**, $\chi^{Model}/\chi_{Bulk}^{Model}$ as a function of $t/\xi_{EX}$ for $\xi_{EX}/\xi_{DM} = 7$ from the Eq. (2). **f**, $\chi^{TB}/\chi_{Bulk}^{TB}$ with respect to $t$ using a minimal TB model.

**Figure 4. Thickness Dependent Orbital-Moment Anisotropy. a**, X-ray absorption spectroscopy (XAS) spectra (solid black line) which is averaged by right- and left-handed circularly polarized X-ray spectra at the Co L$_{2,3}$ edges obtained at 0° for X = Ti and $t_{eff} = 0.6$ nm. **b**, X-ray magnetic circular dichroism (XMCD) spectra (solid black line) corresponding to the difference between right- and left-handed circularly polarized X-ray spectra at the Co L$_{2,3}$ edges obtained at 0° for X = Ti and $t_{eff} = 0.6$ nm. **c**, The magnetic chirality scaled by the SOC constant expressed as $\tilde{\chi}/\xi$ ($\equiv m_\perp/m_\parallel$), obtained from the orbital moments' measurement, as a function of $t_{eff}$ for X = Ta, Ti, Ru, and Pd through an examination of XAS spectra and XMCD spectra. The solid line indicates the best linear fit. The purple arrow represents the dimensionality transition thicknesses $t_C$.

**Acknowledgements**

This work was supported by the National Research Foundation of Korea (NRF) funded by the Ministry of Science and ICT (MSIT) (2022R1A2C2004493, 2020R1A5A1016518, 2022M3H4A1A04096339, 2022R1A4A103134911, 2018R1D1A1B07046980, 2021R1F1A1056444, RS-2024-00334933, RS-2024-00410027) and by the Korea Institute of Science and Technology (KIST) institutional program (2E32251). This study was supported by the Samsung Science and Technology Foundation (SSTF-BA1802-07), Samsung Electronics Co. Ltd. D.H.K. was supported by the POSCO Science Fellowship of the POSCO TJ Park Foundation, the National Research Council of Science and Technology (NST) grant funded by the Ministry of Science and ICT (No. 2N45290 and No. GTL24041-000). S.Y.P. and A.Y.L. were supported by KBSI grant B437200. XAS and XMCD measurements were conducted utilizing the PLS-II 2A beamline.

**Author contributions**

D.Y.K. and D.H.K. conceived the study and D.H.K. supervised the research. S.H.L., J.M., J.Y.C., and B.C.M. were responsible for fabricating the films and devices. D.Y.K. and Y.S.N. conducted the spin-torque measurement experiments, while D.Y.K. and D.H.K. analyzed the symmetric and anti-symmetric exchange interactions. Y.H.K. performed X-ray spectroscopy, and D.H.K. and Y.H.K. conducted orbital moment analysis. A.Y.L. and S.Y.P. carried out experiments related to magnetic properties. K.W.K. provided the theoretical framework and G.G. performed numerical calculations based on a tight-binding model. The manuscript was written by D.H.K. and D.Y.K., with input and comments from K.W.K., G.G., K.J.L., H.Y., and S.B.C. All authors participated in result discussions and manuscript review.

**Additional information**

Correspondence and request for materials should be addressed to H.Y., D.H.K. and S.B.C.

**Competing financial interests**

The authors declare no competing financial interests.

**Data and materials availability**

All data are available in the main text or the Supplementary Information.

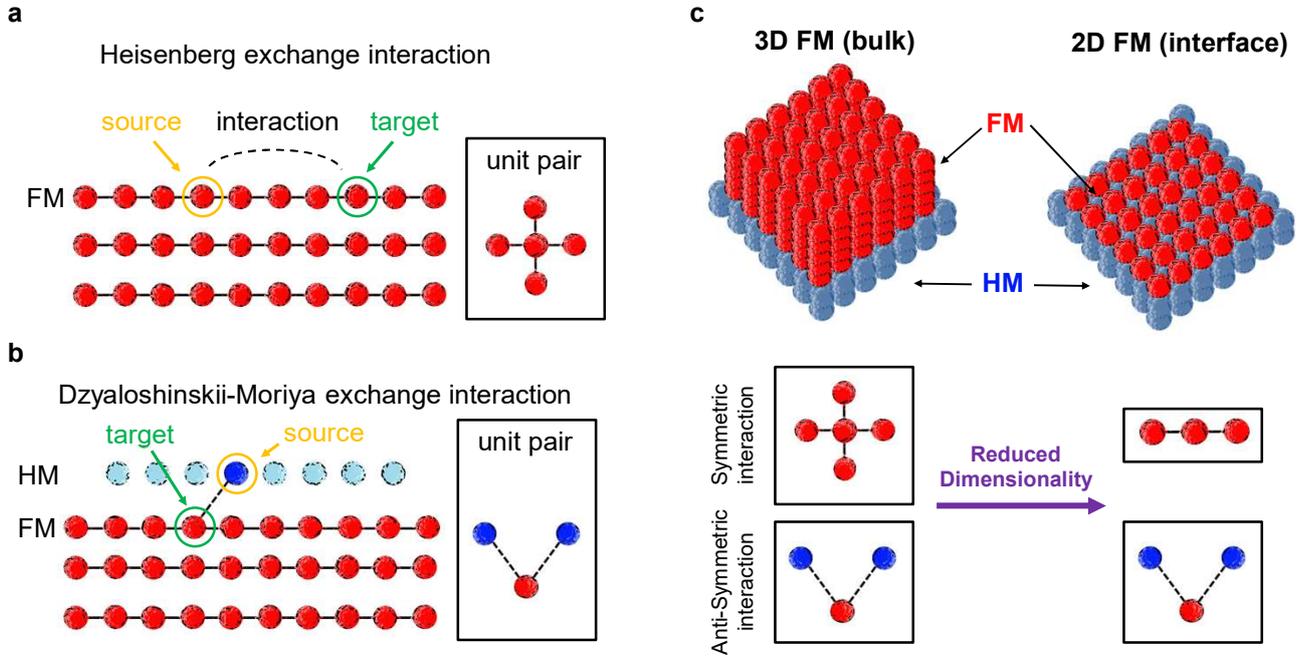

Fig. 1 Kim *et al*.

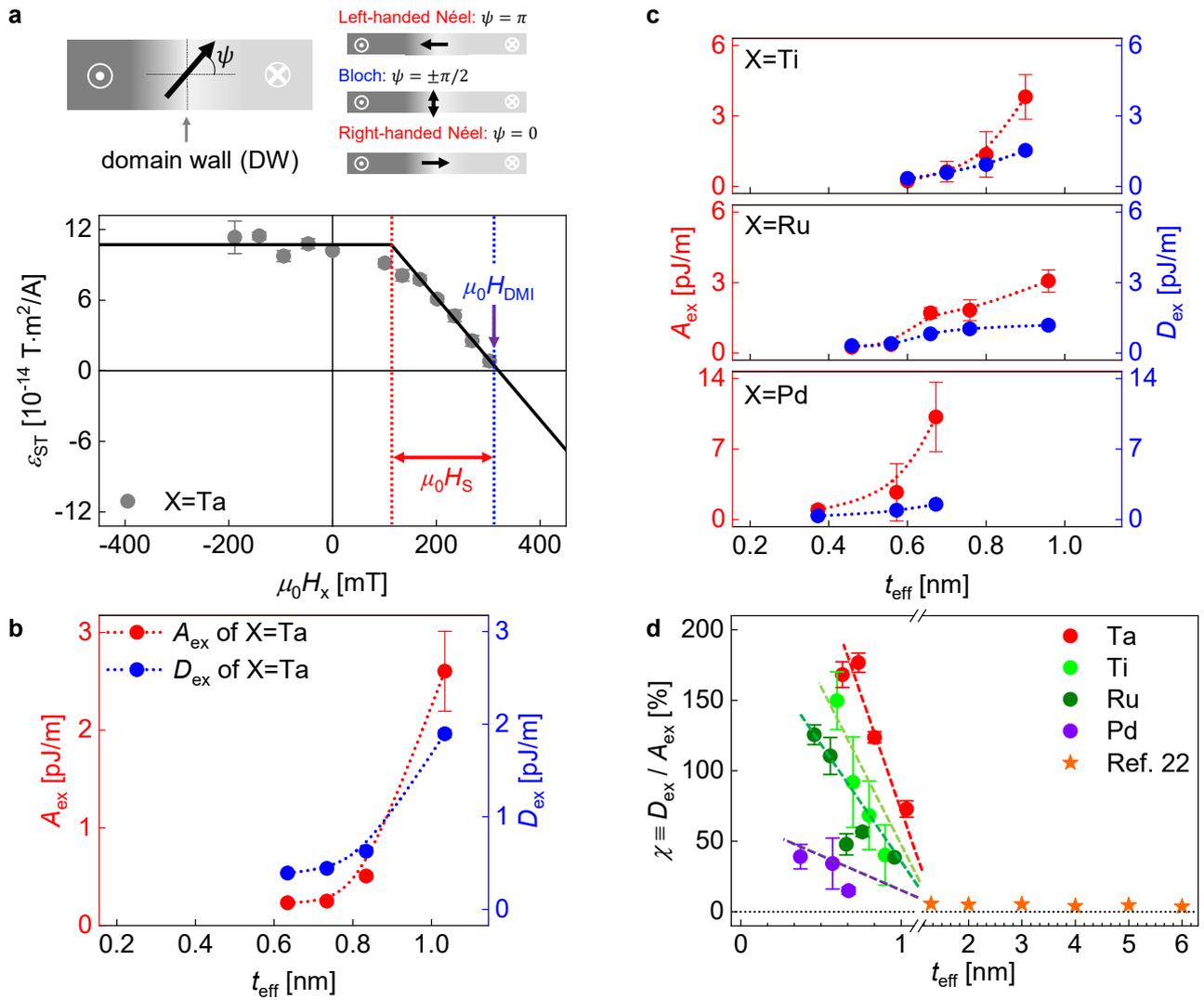

**Fig. 2 Kim *et al.***

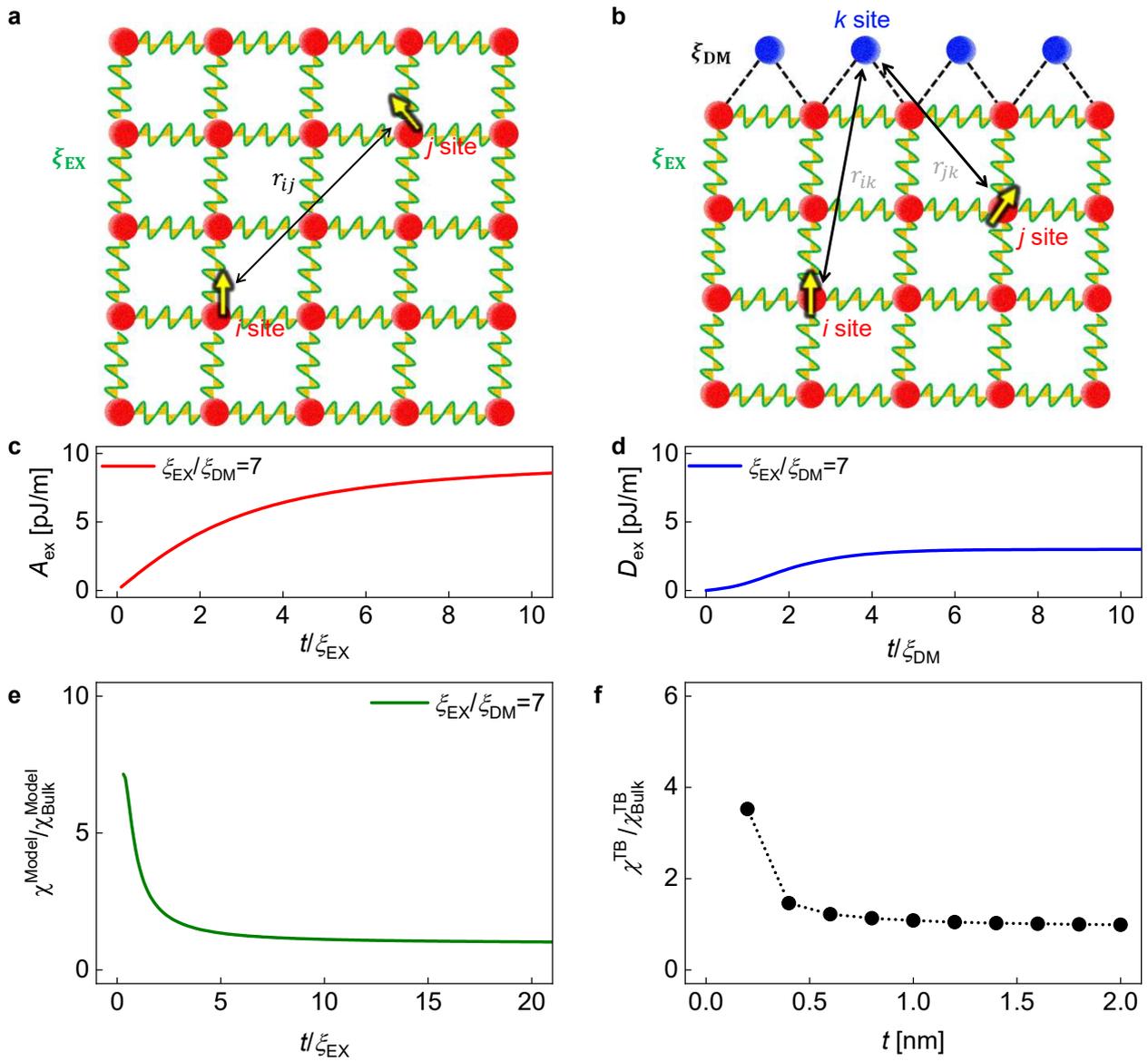

**Fig. 3 Kim *et al*.**

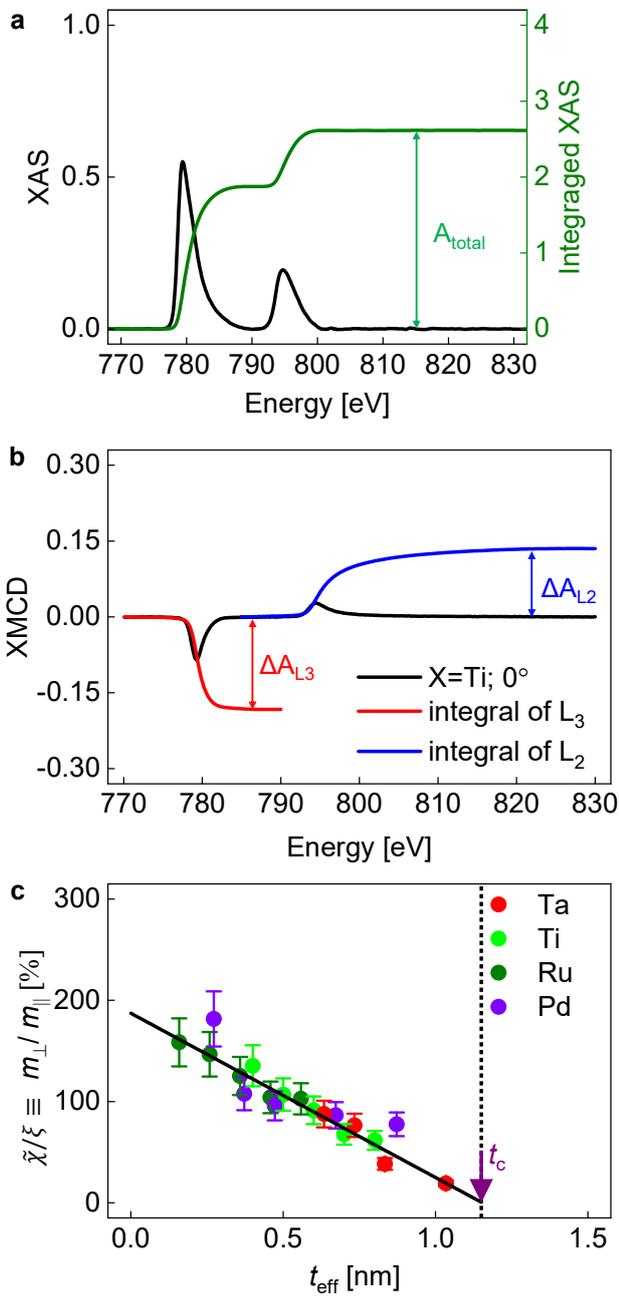

Fig. 4 Kim *et al.*